\documentclass[aps,prd,superscriptaddress,12pt,showpacs,notitlepage]{revtex4}
	\usepackage{graphicx}
	\usepackage{amsmath,amssymb,mathrsfs}

\usepackage{epsf}
\usepackage{epsfig}

\newcommand{\be}{\begin{equation}}
\newcommand{\ee}{\end{equation}}
\newcommand{\bea}{\begin{eqnarray}}
\newcommand{\eea}{\end{eqnarray}}

\newcommand{\bb}{\bibitem}
 
\newcommand{\eqn}{\begin{eqnarray}}
\newcommand{\eqnx}{\end{eqnarray}}

\begin{document}
\title{Topological duality between vortices and planar skyrmions in BPS theories with APD symmetries}
\author{C. Adam}
\affiliation{Departamento de F\'isica de Part\'iculas, Universidad de Santiago de Compostela and Instituto Galego de F\'isica de Altas Enerxias (IGFAE) E-15782 Santiago de Compostela, Spain}
\author{J. Sanchez-Guillen}
\affiliation{Departamento de F\'isica de Part\'iculas, Universidad de Santiago de Compostela and Instituto Galego de F\'isica de Altas Enerxias (IGFAE) E-15782 Santiago de Compostela, Spain}
\author{A. Wereszczynski}
\affiliation{Institute of Physics,  Jagiellonian University,
Reymonta 4, Krak\'{o}w, Poland}
\author{W. J. Zakrzewski}
\affiliation{Department of Mathematical Sciences, University of Durham, Durham DH1 3LE, U.K.}


\begin{abstract}
The BPS baby Skyrme models are submodels of baby Skyrme models, where the nonlinear sigma model term is suppressed. They have skyrmion solutions saturating a BPS bound, and the corresponding static energy functional is invariant under area-preserving diffeomorphisms (APDs).  
Here we show that the solitons in the BPS baby Skyrme model, which carry a nontrivial topological charge $Q_{b} \in \pi_2(S^2)$ (a winding number), are dual to vortices in a BPS vortex model with a topological charge $Q _{v}\in \pi_1(S^1)$ (a vortex number), in the sense that there is a map between the BPS solutions of the two models. The corresponding energy densities of the BPS solutions of the two models are identical. 
A further consequence of the duality is that the dual BPS vortex models inherit the BPS property and the infinitely many symmetries (APDs) of the BPS baby Skyrme models. Finally, we demonstrate that the same topological duality continues to hold for the U(1) gauged versions of the models. 
\end{abstract}

\maketitle 
\section{Introduction}
It is well known  that there is an intimate relation between $SU(2)$ Yang-Mills-Higgs monopoles (t'Hooft-Polyakov monopoles) and skyrmions. Indeed, topologically nontrivial solutions of the Skyrme model look similar to the BPS monopoles in the sense that the energy densities of the corresponding solutions with the same topological charges possess exactly the same symmetries. These solutions are not identical, but there is an approximate qualitative agreement in their corresponding energy density distributions  \cite{prasad}-\cite{skyrmion}. The correct explanation of this fact is given by the use of the rational maps \cite{hough} i.e, holomorphic functions $R(z)=p(z)/q(z)\; : \; S^2 \rightarrow S^2$, where $p,q$ are polynomials without common roots, whose degrees ${\rm deg}(p), {\rm deg}(q)$ define the degree $N$ of the rational map $R$ via $N={\rm max}({\rm deg}(p),{\rm deg}(q))$. As shown in \cite{jarvi} there is a one-to-one correspondence between rational maps of degree $N$ and $N$-monopoles. In particular, any rational map can be derived from a monopole field configuration and,  conversely, each monopole defines uniquely (up to a Mobius transformation) a rational map. On the other hand, the rational maps are behind the famous {\it Rational Map Ansatz} which has proved to be an extremely powerful tool for the construction of approximate soliton solutions in the {\it massless} Skyrme model \cite{hough}.  
\\

This close correspondence no longer seems to work when a non-zero pion mass is included. Obviously, this case is more relevant from the physics perspective and, therefore, any progress in the understanding of geometric properties of skyrmions in this model is important (see, for example, \cite{atiyah}, where skyrmions with massless pions in hyperbolic space are used to approximate skyrmions with massive pions in flat space). 
\\

Recently, inspired by the phenomenological deficiencies of the original Skyrme model, two BPS Skyrme like theories have been proposed. Here, BPS denotes the fact that the solutions 
of these models are self-dual, solving a kind of Bogomolny equations which, hence, leads to a linear energy-topological charge relation. The first model is a {\it conformal} BPS model \cite{conf bps} with massless pions (no potential term) and an infinite tower of KK mesonic fields derived by a dimensional reduction from a higher dimensional Yang-Mills theory. The second model is a {\it volume preserving diffeomorphism} (VPD) BPS model \cite{vpd bps1}, \cite{vpd bps2} with formally infinitely heavy perturbative pions (see also \cite{arthur}, \cite{bps-Sk BM}). Of course, QCD is neither conformal nor VPD invariant, but it has been argued that a BPS model might be a proper starting point for the construction of a correct low energy effective action. In addition, due to the extremely large symmetry groups of the  BPS models it is much easier to find their
classical solutions. Note that, as the latter model contains a potential term for the $SU(2)$ chiral field, it may also provide an analytical insight into Skyrme type theories with a non-trivial potential, where, as we have pointed out, no reliable approximation approach (except for the full numerical simulations) has as yet been found. 
\\

The present paper is devoted to the further investigation of properties of VPD BPS models. In particular, it would be very interesting to study whether there is any correspondence between skyrmions in the VPD BPS model (topological solitons with charges $Q \in \pi_3(S^3)$) and monopoles (solitons with charges $Q \in \pi_2(S^2)$). 
\\

As usual, we want to begin our investigation with a simpler, dimensionally reduced Skyrme type theory, i.e., the baby Skyrme model which is a planar version of the original Skyrme model \cite{baby}. It is known that there also exists a (2+1) counterpart of the VPD BPS Skyrme model called {\it area preserving diffeomorphisms} BPS baby model \cite{bps-baby GP}, \cite{bps-baby}, \cite{bps-baby Sp1}, \cite{bps-baby Z} which consists of two parts: the topological current squared and a potential
\begin{equation}
\mathcal{S}_{BPS \; baby} =\int d^3x \; \left(   -\frac{\lambda^2}{4} (\partial_\mu \vec{\phi} \times \partial_\nu \vec{\phi})^2-\mu^2 V (\vec{n} \cdot \vec{\phi}) \right) .
\end{equation}
In the planar model, instead of the chiral Skyrme field $U \in SU(2)$, one deals with a three component unit vector $\vec{\phi} \in S^2$. Hence, static baby skyrmions are maps from compactified two dimensional base space $\mathbb{R}^2 \cup \{ \infty \} \cong S^2$ into the target space $S^2$ and are, therefore, classified by the corresponding winding number $Q_b \in \pi_2(S^2)$. Obviously, a lower-dimensional counterpart of (magnetic) monopoles are (magnetic) vortices with topological charge $Q_v \in \pi_1 (S^1)$. The precise aim of the present work is, therefore, to show that the BPS  baby skyrmions are related to some vortices. In order to accomplish this  one needs to construct a model in (2+1) dimensions which possesses vortex solutions. As vortex solutions should reflex properties of the BPS baby skyrmions, one is naturally led to consider vortex models which possess the APD symmetry, as well.  Concretely, in Section II we construct such a vortex model and prove that it is, in fact, dual to the BPS baby Skyrme model in the sense that there exists a field transformation mapping it into the latter. Further, we show that the vortex model has a BPS bound and its static vortices saturate this bound. In Section III we repeat the same construction for the models coupled to electromagnetism in the standard way. Section IV contains our conclusions.  
\section{Vortices with the APD Symmetry}
For models with the standard kinetic term, finite energy vortices can be obtained only if an $U(1)$ gauge field is added as an extra independent field to the Lagrangian. The reason  is that the standard kinetic term (i.e., the quadratic sigma model term) produces infinite energy if the vortex boundary conditions are assumed. Let us demonstrate this for the complex scalar case. The kinetic part of the static energy integral is
$$ \int d^2 x \nabla u \nabla \bar{u} = \int r dr \left( f_r^2 +\frac{n^2}{r^2} f^2 \right) $$ 
where the axially symmetric ansatz has been used $u=f(r)e^{in\phi}$. A non-zero value of the integer $n$, together with a proper boundary condition for the profile function $f$, i.e., $f(r=\infty)=f_0 \neq 0$ guarantee a nontrivial topological charge. Unfortunately, this leads to a logarithmical divergency in the energy integral due to the angular term. This divergency can be removed by adding a gauge field, whose angular $A_\phi$ component cancels the non vanishing angular part of the kinetic term. As a consequence, we arrive, e.g., at the well-known Abelian Higgs model \cite{ab higgs}.   
\\

In theories with generalized kinetic terms, where the standard quadratic kinetic term is absent, this argument no longer holds. It is, therefore, possible to construct a planar model based only on a complex field degree of freedom which does support finite energy vortices without any need for a gauge field. The omission of the ``usual'' quadratic kinetic term might be seen as a drastic modification of a theory. We point out that the kinetic terms we consider in this paper are still quadratic in first time derivatives, such that a standard Hamiltonian may be derived. The non-negative coefficients of these derivative terms can be zero for some field configurations (specifically, they vanish for vacuum configurations), therefore the Cauchy problem is not well defined for the corresponding initial data. However, the model is thus an effective model which possesses extra symmetries which allow us to say much more about the properties 
of its static field configurations. The complete model will possess the ``usual" kinetic terms (either added explicitly or generated, e.g., by the quantum corrections) but we hope, most other 
properties will not be altered too much.  From a more physical perspective, depending on the field theory under consideration, omitting the quadratic kinetic term may correspond to an approximation which is quite reliable in a nonperturbative regime (where the energy density is rather large), whereas it does not reproduce the behaviour near the vacuum (linear fluctuations about the vacuum are completely suppressed).             
\subsection{Higgs potential}
The model we want to discuss in this section is defined on a (2+1) dimensional Minkowski space-time manifold, with the Lagrangian
\begin{equation}
\mathcal{L}_{BPS \; vortex}=\lambda^2 \mathcal{L}_4+\mathcal{L}_0 \label{bps vortex}
\end{equation}
which consists of two parts:  a fourth derivative term (a Skyrme-like term)
\begin{equation}
\mathcal{L}_4=-(u_\mu \bar{u}^\mu)^2 + u_\mu^2\bar{u}_\nu^2 = -K_\mu u^\mu ,
\end{equation}
where we have introduced 
\begin{equation}
K_\mu= (u_\nu \bar{u}^\nu) \bar{u}_\mu - \bar{u}_\nu^2 u_\mu ,
\end{equation}
and a non-derivative part, i.e., a potential
\begin{equation}
\mathcal{L}_0= -V(u\bar{u}) .
\end{equation}
It is straightforward to see that such a model circumvents the usual Derrick argument against the existence of the static finite energy solutions and, therefore, may support solitons. As the model is invariant under the area preserving diffeomorphisms (precisely speaking its static energy integral is invariant under area preserving diffeomorphisms of the base space \cite{bps-baby Z}, \cite{vpd bps2}) and, as is shown below, has solutions which obey a Bogomolny equation, it is natural to call this model the {\it APD BPS vortex} model. 
\\  

The Euler-Lagrange field equation  corresponding to (\ref{bps vortex}) takes the form
\begin{equation}
\partial_\mu K^\mu - \frac{1}{2\lambda^2} V_u=0,
\end{equation}
and in the static case we find that it reduces to 
\begin{equation}
\nabla \vec{K} - \frac{1}{2\lambda^2} V_u=0, \;\;\;\; \vec{K}=(\nabla u \nabla \bar{u}) \nabla \bar{u} - (\nabla \bar{u})^2 \nabla u .
\end{equation}
For the moment, we  select the potential of the Abelian Higgs model 
\begin{equation} \label{Higgs-pot}
V=\frac{\mu^2}{4} \left( 1-u\bar{u} \right)^2
\end{equation}
(we choose the vacuum at $|u|=1$, which can always be achieved by rescaling $u$ by a real constant and by a corresponding rescaling of the coupling constants).
Next, we consider the usual static axially symmetric ansatz 
\begin{equation}
u(r,\phi) = f(r) e^{in\phi}
\label{ans}
\end{equation}
where $n \in Z$ is the winding number and $f$ is a profile function whose form remains to be determined. Then, we arrive at the following equation for the profile function $f$,
\begin{equation}
f \frac{1}{r} \partial_r \left( \frac{ff_r}{r} \right) + f \frac{\mu^2}{8n^2\lambda^2} \left( 1 -f^2\right)=0 , \label{prof f}
\end{equation}
where one has to assume the obvious vortex-like boundary conditions leading to the nontrivial topological charge
\begin{equation}
f(r=0)=0, \;\;\;\;\; f(r=\infty)=1.
\end{equation}
This equation can be further simplified by introducing a new target space variable
\begin{equation}
h=1- f^2
\end{equation}
together with a new base space coordinate 
\begin{equation}
x=r^2/2 .
\end{equation}
Then, we get (we omit the overall factor $f$)
\begin{equation}
h_{xx} - \frac{\mu^2}{4n^2\lambda^2} h=0,
\end{equation}
and the boundary conditions become
\begin{equation}
h(x=0)=1, \;\;\;\;\; h(x=\infty)=0.
\end{equation}

A  topologically nontrivial solution is then given by
\begin{equation}
h=e^{-\frac{\mu}{2n \lambda}x} \;\;\; \Rightarrow \;\;\; f=\sqrt{1-e^{-\frac{\mu}{4n\lambda}r^2}},
\end{equation}
and the total energy of this solution becomes
\begin{eqnarray}
E&=& \int d^2x \left( \lambda^2[(\nabla u \nabla \bar{u})^2  - (\nabla \bar{u})^2 (\nabla u)^2] +V \right) \nonumber \\
&=& \int d^2x \left( 4n^2\lambda^2\frac{f^2f_r^2}{r^2} +V \right) =  2\pi \int_0^{\infty}dx  (4n^2\lambda^2 f^2f_x^2+V)  \nonumber \\ 
&=& 2\pi n^2 \lambda^2 \int_0^\infty dx  \left(  h_x^2 +  \frac{\mu^2}{4n^2\lambda^2} h^2 \right) =  \pi \mu^2 \int_0^\infty dx   h^2 = \pi \mu   \lambda n . \label{Higgs-energy}
\end{eqnarray}
Hence, $E$ is proportional to the topological charge of the soliton solution which demonstrates the BPS nature of the solutions.
\\

One should note that the profile function equation (\ref{prof f}) has a well defined solution for more general boundary conditions i.e., $f(r=0)=f_0$, where $f_0$ is an arbitrary constant. This is different from the standard vortex models where such a condition is excluded if one imposes regularity at the origin. Here, as far as the field equation is considered, the profile equation does not require such a restriction. However, from the point of view of the $h$ function this leads to a multiplication of the original solution by a constant $h_0=\sqrt{1-\frac{\mu^2}{m^2} f_0^2}$. This constant also shows up in the energy and would result in a continuos spectrum of finite energy solutions in a fixed topological sector. This would then be  not acceptable, as one could find solutions with arbitrarily small energy. Fortunately, such solutions are excluded if we require to have single-valued configurations at the origin, and the previously assumed boundary condition $f(0)=0$ is the only acceptable one to guarantee this.  
\subsection{generalized Higgs potentials}
The results obtained above may be easily extended to the case of a family of generalized Higgs potentials 
\begin{equation}
V=\frac{\mu^2}{4} \left( 1-u\bar{u} \right)^\alpha,
\end{equation}
where the parameter $\alpha \geq 1$. Assuming the same ansatz and performing the same change of the target and base space variables we arrive at the following equation for $h$,  
\begin{equation}
h_{xx} - \frac{\alpha \mu^2}{8n^2\lambda^2}  h^{\alpha-1}=0
\end{equation}
with a solution (for $\alpha>2$)
\begin{equation}
h(x)=\left(\frac{x_0}{x+x_0} \right)^{\frac{2}{\alpha-2}}, \;\;\;\;\;\; x_0=\frac{4n\lambda}{\mu (\alpha-2)} 
\end{equation}
which describes a power-like localized vortex. Localization becomes weaker with growing $\alpha$. For $\alpha \in [1,2)$ we find compact vortices i.e., solitons for which
the field takes the vacuum value at a finite distance
\begin{equation}
h(x)= \left\{
\begin{array}{cc}
\left( 1-\frac{x}{x_0}\right)^{\frac{2}{2-\alpha}} & x \in [0,x_0]
\\
 & \\
 0 & x \geq x_0
\end{array}
\right. \;\;\;\;\;\;\; x_0=\frac{4n\lambda}{\mu (2-\alpha)} .
\end{equation}
Specifically, for $\alpha=1$ we get the standard signum-Gordon compacton with the parabolic approach to the vacuum \cite{arodz}. We remark that these compactons always are genuine minimizers of the corresponding variational problem.
\subsection{Bogomolny equation and BPS bound}
Now we prove that the static energy of the model has a topological bound which is saturated by solutions of a certain first order equation usually referred to as the BPS equation. Let us start by completing a square in the total energy, as usual,   
\begin{eqnarray}
E_{BPS\; vortex} &=& \int d^2x \left( \lambda^2[(\nabla u \nabla \bar{u})^2  - (\nabla \bar{u})^2 (\nabla u)^2] +V \right)
\nonumber \\ &=&
\int d^2x \left( i \lambda  \epsilon_{ij} \nabla_i u \nabla_j \bar{u}  \pm  \sqrt{V} \right)^2 \mp  2 i \lambda \int d^2x \epsilon_{ij} \nabla_i u \nabla_j \bar{u} \sqrt{V}
\end{eqnarray}
and giving us
\begin{equation}
E_{BPS\; vortex} \geq B_{BPS} \equiv \vert 2i \lambda \int d^2x  \epsilon_{ij} \nabla_i u \nabla_j \bar{u} \sqrt{V} \vert
\end{equation}
with the equality holding if and only if the following BPS equation is statisfied:
\begin{equation}
\lambda \epsilon_{ij} \nabla_i u \nabla_j \bar{u}=\pm i \sqrt{V} .
\end{equation}

One can check that, when the axially symmetric ansatz is inserted into this BPS equations, it agrees with the first integral of the static Euler-Lagrange equation for the profile function. We still have to demonstrate that the bound is topological, i.e., equal to a universal constant times the vortex number $n$. This may be seen easily, starting from the observation that the base space two-form $d^2 x \epsilon_{ij} \nabla_i u \nabla_i \bar{u} \sqrt{V}$ is, in fact, the pullback of a two-form on the target space, and the base space integral may, therefore, be replaced by a target space integral. In our case the target space is just $\mathbb{R}^2$ and, therefore, the two-form on the target space is exact (a total derivative), but it gives a nontrivial result, nevertheless, because of the nontrivial boundary conditions imposed on the target space coordinates (the field $u$) by the Higgs potential. Indeed, introducing the real cartesian target space coordinates $X,Y$ via $u=X+iY$ we find that $\epsilon_{ij} \nabla_i u \nabla_j \bar{u} = -2i \epsilon_{ij} \nabla_i X \nabla_j Y  $.  So, for the BPS bound $B_{BPS}$ we have
\begin{equation}
B_{BPS} = 4\lambda \int d^2 x \epsilon_{ij} \nabla_i X \nabla_j Y \sqrt{V} = 4\lambda n \int d^2 X \sqrt{V(X^2 + Y^2)}
= 4\pi \lambda n \int_0^{1} dt \sqrt{V(t)}
\end{equation}
where we have introduced polar coordinates $t= X^2  + Y^2$ and $\Phi$ in the target space. Furthermore, we have assumed that the Higgs type potentials take their vacuum values at $t=1$, $V(t=1)=0$. The factor $n$ in front of the target space integral is due to the fact  that, for a field configuration with vortex number $n$, the target space is covered $n$ times while the base space is covered once. Obviously, the bound depends only on the potential and the coupling constants of the model under consideration, as well as on the vortex number, as is required for a topological BPS bound. For the specific Higgs potential (\ref{Higgs-pot}), we easily recover the energy (\ref{Higgs-energy}) from the above expression.
\subsection{Duality between the BPS vortices and BPS baby Skyrmions}
For convenience, we start with the BPS baby Skyrme model in its $CP^1$ formulation 
\begin{equation}
\mathcal{L}_{BPS \; baby}= - \lambda^2_b \frac{K_\mu u^\mu}{(1+|u|^2)^4} - \frac{\mu^2_b}{4} \left( \frac{|u|^2}{1+|u|^2}\right)^\alpha , \label{bps baby}
\end{equation}
which differs from the quartic vortex model by a target space factor multiplying the fourth derivative part and by a different family of potentials (the so-called old baby Skyrme potentials). As was pointed out before, these models are different also at a deeper level, as they support solitonic solutions with a completely distinct topological nature. The baby skyrmions are maps from compactified two dimensional base space $\mathbb{R}^2 \cup \{ \infty \} \cong S^2$ into the target space, which is also $S^2$, and are, therefore, classified by the winding number $Q \in \pi_2(S^2)$. For the radially symmetric ansatz this implies that the profile function $f_{b}$ must cover the whole semi-line. Usually, one choses $f(r=0)=\infty$ and $f(r=R)=0$, where $R$ can be finite (compactons) or infinite (standard baby skyrmions with infinite tails).      
\\

Our main observation is that the baby skyrmions of the BPS baby Skyrme model (\ref{bps baby}) and the vortices of the BPS vortex model (\ref{bps vortex}) are related by a non-holomorphic transformation. Such a map influences the corresponding boundary conditions and, therefore, transforms the topological properties of one model into the other. In a sense, it demonstrates that the vortices and baby skyrmions in these BPS models are, in fact, dual objects governed by the same "master" energy density. 

\vspace*{0.2cm}

\noindent {\bf Proposition}: There is a one-to-one correspondence between solutions of the APD BPS vortex model $u_v$ and solutions of the baby BPS model $u_b$. The solutions are related via a non-holomorphic transformation: 
\begin{equation}
f^2_v (x_\mu)=  \frac{1}{1+f_b^2(x_\mu)}, \;\;\;\; \Phi_v (x_\mu)=\Phi_b(x_\mu),
\end{equation}
where the real functions $f_v, f_b$ and $\Phi_v, \Phi_v$ are defined as
 \begin{equation}
 u_v(x_\mu) = f_v(x_\mu) e^{i\Phi_v(x_\mu)}, \;\;\;\;\; u_b(x_\mu) = f_b(x_\mu) e^{i\Phi_b(x_\mu)} .
\end{equation}
Moreover, the coupling constants of the models remain unchanged,
\begin{equation}
\lambda_b=\lambda , \;\;\; \mu_b=\mu .
\end{equation}
\noindent {\bf Proof}: It suffices to show that the corresponding Lagrange densities are connected by means of this transformation. So, 
\begin{eqnarray}
- \mathcal{L}_{BPS\; vortex}= 
 \lambda^2 [(u_\mu \bar{u}^\mu)^2-u_\mu^2\bar{u}_\nu^2 ] + \frac{\mu^2}{4} \left( 1 - |u|^2 \right)^\alpha = 
 \\
 4\lambda^2 f_v^2 [(\partial_\mu f_v)^2 (\partial_\nu \Phi_v)^2  - (\partial_\mu f_v \partial^\mu \Phi_v)^2] + \frac{\mu^2}{4} \left( 1 - f_v^2 \right)^\alpha =
   \\
 4 \lambda^2  \frac{f_b^2}{(1+f_b^2)^4} [(\partial_\mu f_b)^2 (\partial_\nu \Phi_b)^2  - (\partial_\mu f_b \partial^\mu \Phi_b)^2] + \frac{\mu^2}{4} \left( \frac{f_b^2}{1+f_b^2} \right)^\alpha =\\
4 \lambda^2_b  \frac{f_b^2}{(1+f_b^2)^4} [(\partial_\mu f_b)^2 (\partial_\nu \Phi_b)^2  - (\partial_\mu f_b \partial^\mu \Phi_b)^2] + \frac{\mu^2_b}{4}  \left( \frac{f_b^2}{1+f_b^2} \right)^\alpha =  - \mathcal{L}_{BPS \; baby} .
\end{eqnarray}
For a general potential $V_b(f^2_v)$ of the BPS vortex model, the potential $V_b$ of the dual model is simply given by
\begin{equation} \label{vb-pot}
V_b (f^2 _b) = V_v \left(\frac{1}{1+f_b^2 }\right)
\end{equation}
where the vacuum of $V_v$ at $f_v =1$ transforms into the corresponding vacuum of $V_b$ at $f_b =0$.

\vspace*{0.2cm}

\noindent {\bf Observation 1}: The static vortices of the APD BPS vortex model are transformed into the baby skyrmions of the BPS baby Skyrme model with the equality of the corresponding topological charges
\begin{equation}
Q_{b}=Q_{v}
\end{equation}
\noindent {\bf Proof}: 
In the case of the axially symmetric static solutions found above, the transformation maps the profile function of a vortex solution of the BPS vortex model into the profile function of a baby skyrmion of the BPS baby model
\begin{equation}
f^2_{v}(r)= \frac{1}{1+f^2_{b}(r)} \label{dual static}
\end{equation}
while the remaining angular dependent parts of the complex fields in these models remain unchanged, i.e., 
\begin{equation}
n_{baby}=n_{vortex}
\end{equation}
As (\ref{dual static}) maps the proper baby skyrmion boundary conditions into vortex boundary conditions, this proves the claimed result.  

\vspace*{0.2cm}

\noindent {\bf Observation 2}: The BPS sector (equation) of one model is mapped into the BPS sector (equation) of the other one. 
\\
\noindent {\bf Proof}: By a simple application of the transformation map.

\vspace*{0.2cm}

\noindent {\bf Observation 3}: Both models (in the static version and using the ansatz (\ref{ans})) possess the same {\it master} dimensionally reduced energy integral 
\begin{equation}
E=2\pi \int_0^\infty dx \left( n^2\lambda^2 h_x^2 +\mu^2 h^\alpha \right) 
\end{equation}

\noindent {\bf Proof}: 
Let us write the static energy for the BPS baby model for the ansatz (\ref{ans}) 
\begin{eqnarray}
E_{BPS \; baby}=2\pi \int_0^\infty r dr  \left( 4n^2\lambda^2_b \frac{f^2f_r^2}{r^2(1+f^2)^4} +\mu^2_b \left( \frac{f^2}{1+f^2}\right)^\alpha \right)
\\
= 2\pi \int_0^\infty dx \left( n^2\lambda^2_b h_x^2 +\mu^2_b h^\alpha \right), 
\end{eqnarray}
where we have introduced
\begin{equation}
h=1-\frac{1}{1+f^2_b}
\end{equation}
Up to an immaterial multiplicative factors this expression is exactly the same energy density (in terms of the $h$ profile function) as for the BPS vertex model with the family of generalized Higgs potentials. Moreover, $h$ obeys exactly the same boundary conditions as its counterpart defined for the vortex case. Namely, $h(0)=1$ and $h(R)=0$.  

\vspace*{0.2cm}

\noindent {\bf Observation 4}: Time-dependent spinning configurations rotate with the same frequency
\\ 
\noindent {\bf Proof}:  The spinning solutions are obtained from the following ansatz  
\begin{equation}
u= f(r) e^{in \phi +i \omega t}
\end{equation}
Hence, using the transformation map we get $(n,\omega)_{vortex}=(n,\omega)_{baby}$. 

\vspace*{0.2cm}

\noindent All these results show that both models are in fact dual to each other. They describe exactly the same physics (identical energy densities and symmetries) but by means of two different topological objects. Therefore we call this duality a {\it topological duality}. Such a duality is rather unusual, as it relates two distinct topological charges. Typical examples of dualities transform a topological charge of one model into a Noether charge of its dual counterpart (see eg. T-duality \cite{T duality} and Montonen-Olive duality \cite{olive} between electric and magnetic charges).  
  \\

Rather surprisingly, we have found that the link between baby skyrmions and vortices in the APD BPS models is much more intimate than in the case of the usual skyrmions and monopoles. Here, they are not only qualitatively similar, they are essentially identical.   
\\
Let us again stress that the vortex solutions are not of a magnetic type, as we have  not introduced any gauge field. So, the duality connects BPS baby skyrmions with "complex scalar" vortices without any magnetic flux. However, it is possible to relate such skyrmions with proper magnetic vortices of an Abelian-Higgs type model. The only thing we need to do is to gauge the APD BPS models by the minimal coupling with the Maxwell field. 
\section{Abelian-Higgs Model with the APD Symmetry}
\subsection{Higgs potential}
The APD BPS vortex model minimally coupled to the Maxwell field is simply given by the following Lagrange density
\begin{equation}
\mathcal{L}_{gauged \;BPS \;vortex}=-\lambda^2 [ (D_\mu u D^\mu \bar{u})^2- (D_\mu u)^2 (D_\nu \bar{u})^2] - V(u\bar{u}) - \frac{1}{4g^2} F_{\mu \nu}^2,
\end{equation}
where the covariant derivative is given by
\begin{equation}
D_\mu u = u_\mu -i A_\mu u .
\end{equation}
This is just the APD version of the Abelian Higgs model where, for the moment, we assume the Higgs (Mexican hat) potential. Then, the equations of motion become 
\begin{equation}
\bar{D}_\mu \mathcal{K}^\mu  - \frac{1}{2} V_u=0,
\end{equation}
where
\begin{equation}
\mathcal{K}^\mu = (D_\nu u D^\nu \bar{u}) D^\mu \bar{u} -  (D_\nu \bar{u})^2 D^\mu u, \;\;\;\; \bar{D}_\mu \mathcal{K}^\mu= (\partial_\mu +i A_\mu) \mathcal{K}^\mu
\end{equation}
and 
\begin{equation}
\frac{1}{g^2} \partial_\mu F^{\mu \nu}- 2ie \left[ (D_\mu u D^\mu \bar{u}) ( \bar{u}D^\nu u-u D^\nu \bar{u}) - (D_\mu u)^2 \bar{u}D^\nu \bar{u} - (D_\mu \bar{u})^2 u D^\nu u \right]=0 .
\end{equation}
Again, we assume the static ansatz with  
\begin{equation}
A_0=A_r=0, \;\;\; A_\phi=na(r) .
\end{equation}
In this case the matter equations give us
\begin{equation}
D_i u D_i \bar{u}= f_r^2+\frac{n^2f^2}{r^2} (1-a)^2, \;\;\;  (D_i \bar{u})^2 = \left(f_r^2-\frac{n^2f^2}{r^2} (1-a)^2\right) e^{-2in\phi} ,
\end{equation}
and the static field equation becomes
\begin{eqnarray}
0=\partial_x [f^2 f_x (1-a)^2] -f_x^2f(1-a)+af_x^2f(1-a) +\frac{\mu^2}{8n^2\lambda^2} f \left( 1 -f^2\right)
=\\ \partial_x [f^2 f_x (1-a)^2] -f_x^2f(1-a)^2 +\frac{\mu^2}{8n^2\lambda^2} f \left( 1 -f^2\right)
= \\f \left[ \partial_x [f f_x (1-a)^2] +\frac{\mu^2}{8n^2\lambda^2}  \left( 1 -f^2\right) \right] ,
\end{eqnarray}
or in terms of the $h$ function
 \begin{equation}
f \left[ \partial_x [h_x (1-a)^2] -\frac{\mu^2}{4n^2\lambda^2}  h \right]=0 .
\end{equation}

Similarly, one can simplify the Maxwell equation (with a source given by the complex scalar field) obtaining 
\begin{equation}
a_{rr}-\frac{a_r}{r}= - 8 g^2 f^2f_r^2(1-a)
\end{equation}
i.e.,
\begin{equation}
a_{xx}=-2g^2 h_x^2(1-a) .
\end{equation}
Let us also shift $a \rightarrow -a$, which corresponds to changing the sign of the electric charge. In this case we arrive at the following set of equations 
 \begin{equation}
\partial_x [h_x (1+a)^2] -\frac{\mu^2}{4n^2\lambda^2}  h=0
\end{equation}
\begin{equation}
a_{xx}=2g^2 h_x^2(1+a) ,
\end{equation}
which are exactly equal to the static field equations for the gauged BPS baby Skyrme model with potential $V=4h^2$ (up to a trivial numerical factor) - see section V.C. of Ref. \cite{bps-Sk1}. \\

It is straightforward to generalize the above formulas to the case of the one-parameter family of potentials introduced above. The corresponding static equations then are given by
 \begin{equation}
\partial_x [h_x (1+a)^2] -\frac{\alpha \mu^2}{8n^2\lambda^2}  h^{\alpha-1}=0
\end{equation}
\begin{equation}
a_{xx}=2g^2 h_x^2(1+a)
\end{equation}
and we note that, obviously, only the first equation has been modified. 
\subsection{Duality between vortices and BPS baby Skyrmions}
Here, we establish a duality between vortices in the gauged BPS vortex model and baby skyrmions in the gauged BPS baby Skyrme model, whose Lagrange density is  
\begin{equation}
\mathcal{L}_{gauged \;BPS \;baby}=-\lambda^2 \frac{ (D_\mu u D^\mu \bar{u})^2- (D_\mu u)^2 (D_\nu \bar{u})^2]}{(1+|u|^2)^4} - \mu^2 \left( \frac{|u|^2}{1+|u|^2} \right)^\alpha - \frac{1}{4g^2} F_{\mu \nu}^2 ,
\end{equation}
where we have chosen a specific form of one vacuum potentials. Assuming a gauged extended version of the duality map
\begin{equation}
f^2_v (x_\mu)= \frac{m^2}{\mu^2} \frac{1}{1+f_b^2(x_\mu)}, \;\;\;\; \Phi_v (x_\mu)=\Phi_b(x_\mu), \;\;\;\; A^v_\mu (x_\mu)=A^b_\mu (x_\mu)
\end{equation}
we get
\begin{eqnarray}
- \mathcal{L}_{gauged \;BPS \;vortex}=\lambda^2 [ (D_\mu u D^\mu \bar{u})^2- (D_\mu u)^2 (D_\nu \bar{u})^2] + V(u\bar{u}) + \frac{1}{4g^2} F_{\mu \nu}^{v \;2} = \\  
 4\lambda^2_v f_v^2 [(\partial_\mu f_v)^2 (\partial_\nu \Phi_v -A^v_\nu)^2  - (\partial_\mu f_v ( \partial^\mu \Phi_v -A_\mu^v)^2] + \frac{\mu^2_v}{4} \left( \frac{m^2}{\mu^2} - f_v^2 \right)^\alpha + \frac{1}{4g^2} F_{\mu \nu}^{v \;2} = \\  
 4 \lambda^2_v \frac{m^4}{\mu^4} \frac{f_b^2}{(1+f_b^2)^4} [(\partial_\mu f_b)^2 (\partial_\nu \Phi_b-A_\nu^b)^2  - (\partial_\mu f_b (\partial^\mu \Phi_b -A_\mu^b)^2] + \frac{\mu^2_v}{4} \left( \frac{m^2}{\mu^2}\right)^\alpha \left( \frac{f_b^2}{1+f_b^2} \right)^\alpha \\ + \frac{1}{4g^2} F_{\mu \nu}^{b \;2}= -  \mathcal{L}_{gauged \; BPS \; baby} .
\end{eqnarray}
The corresponding static energy then becomes (after inserting the ansatz) 
\begin{eqnarray}
E= 2\pi \int_0^\infty rdr \left[ \frac{ 4\lambda^2n^2}{r^2}  \frac{f^2f_r^2(1+a)^2}{(1+f^2)^4} + \mu^2 \left( \frac{f^2}{1+f^2} \right)^\alpha + \frac{n^2}{4g^2} \frac{a_r^2}{r^2} \right]
=\\  2\pi \int_0^\infty dx \left[ \lambda^2n^2  h^2_x + \mu^2 h^\alpha + \frac{n^2}{4g^2} a_x^2 \right],
\end{eqnarray}
where we used $h_{baby}$ to perfom the last step, which exactly agrees with the static energy of the gauged BPS vortex model with the generalized Higgs potential. \\

Let us mention that it is even possible to introduce vortex solutions directly for gauged baby Skyrme models or gauged O(3) sigma models by assuming a symmetry-breaking potential \cite{o3-vortices}. But in these cases the topology of solutions is no longer determined by the topology of the target space manifold but, instead, by the vacuum manifold of the symmetry-breaking potential, so the situation is essentially equivalent to the standard abelian Higgs model.
\subsection{BPS bound and BPS equations}
In this subsection we demonstrate that the gauged BPS vortex model has a BPS bound and BPS equations and, further, that this bound and equations are exactly dual to the corresponding BPS bound and equations of the gauged BPS baby Skyrme model, i.e., they are identical when expressed in terms of the "master" function $h$. First, let us demonstrate that the two static energy densities are identical when expressed in terms of the function $h$. The static energy density of the gauged vortex model is given by
\begin{equation}
  {\cal E}_v = \lambda^2 {\cal Q}_v^2 + V_v + \frac{1}{g^2} B^2,
\end{equation}
where the covariant ``topological density" ${\cal Q}_v$ takes the form
\begin{eqnarray}
{\cal Q}_v &\equiv & i \epsilon_{ij} D_i u D_j \bar u = i\epsilon_{ij} u_i \bar u_j + \epsilon_{ij} A_i \partial_j |u|^2 
\equiv { \bf q}_v + \epsilon_{ij} A_i \partial_j |u|^2 \\
&=& \epsilon_{ij} \partial_i (f_v)^2 (\Phi_j - A_j) = \epsilon_{ij} \partial_i h (A_j - \Phi_j) .
\end{eqnarray}
On the other hand, the static energy density of the gauged BPS baby Skyrme model is given by
\begin{equation}
  {\cal E}_b = \lambda^2 {\cal Q}_b^2 + V_b + \frac{1}{g^2} B^2
\end{equation}
where its covariant ``topological density" ${\cal Q}_b$ takes the form
\begin{eqnarray}
{\cal Q}_b & \equiv & -i (1+u\bar u)^{-2} \epsilon_{ij} D_i u D_j \bar u = -(1+u \bar u)^{-2} \epsilon_{ij} \left( i u_i \bar u_j + \epsilon_{ij} A_i \partial_j |u|^2\right)  \nonumber \\
&=& \epsilon_{ij} \partial_i (1+f_v^2)^{-1}  (\Phi_j - A_j) = \epsilon_{ij} \partial_i h (A_j - \Phi_j) .
\end{eqnarray}
The two topological densities are completely identical when expressed in terms of the ``master function" $h$. The energy densities are, therefore, identical provided that the potentials are the same when expressed as functions of $h$, i.e. $V_v (h) = V_b(h)$. The potentials are different when expressed in terms of their respective fields $u_v$, $u_b$, because the expression of these fields in terms of $h$ are different,
\begin{equation}
h=1-f_v^2 = 1-\frac{1}{1+f_b^2}.
\end{equation}
The potentials are, in fact, just related by Eq. (\ref{vb-pot}). \\

Next we derive the BPS bound and the corresponding BPS equations. We express the fields in terms of $h$, which allows us to treat the two cases simultaneously. Next we consider a suitable non-negative expression, on which we can perform the usual BPS trick, and which is, as we show later,  the energy density minus a topological term. This non-negative expression is given by
\begin{eqnarray}
0 &\le & \lambda^2 ({\cal Q} - w(h))^2 + \frac{1}{g^2}(B+b(h))^2 \nonumber \\
&=& \lambda^2 ({\cal Q}^2 + w^2 ) + \frac{1}{g^2}( B^2 + b^2 ) - 2\lambda^2 {\bf q} w \nonumber \\
&& -2\lambda^2 \epsilon_{ij} w(h) \partial_i h A_j + \frac{2}{g^2} b(h)\epsilon_{ij} \partial_i A_j 
\end{eqnarray}
where $w(h)$ and $b(h)$ are functions of $h$ that are still to be defined. The last two terms in this expression combine into a total derivative if we assume  that these functions are related as
\begin{equation}
b(h) = -g^2 \lambda^2 W(h) , \quad W(h) \equiv \int_0^h dh' w(h') 
\end{equation}
\begin{equation}
 \Rightarrow \quad  
\lambda^2 \epsilon_{ij} w(h) \partial_i h A_j + \frac{1}{g^2} b(h)\epsilon_{ij} \partial_i A_j = -\lambda^2 \epsilon_{ij} \partial_i (WA_j).
\end{equation}
This total derivative does not contribute to the energy and may therefore be omitted, because $W(h)$ is zero at the vacuum value $h=0$ by assumption. The remainder of the non-negative expression may indeed be written as the energy density minus the topological term $2\lambda^2 {\bf q} W_h$ provided that the function $W$ obeys the first order nonlinear ODE (the ``superpotential equation")
\begin{equation} \label{superpot-eq}
\lambda^2 W_h^2 + \lambda^4 g^2 W^2 = V(h).
\end{equation}
Assuming that this is the case  we find for the energy the inequality
\begin{equation}
E=\int d^2 x \left( ({\cal Q} - W_h)^2 + \frac{1}{g^2} (B - g^2 \lambda^2 W)^2 \right) + 2\lambda^2 \vert \int d^2 x {\bf q}W_h \vert
\ge 2\lambda^2 \vert \int d^2 x {\bf q}W_h \vert
\end{equation}
with equality holding if the BPS equations
\begin{equation}
{\cal Q} =W_h \; ,\quad B= g^2 \lambda^2 W
\end{equation}
are satisfied. Finally we note that the bound is topological because
\begin{equation}
\int d^2 x {\bf q} W_h = 2\int d^2 x \epsilon_{ij} \partial_i X \partial_j Y W_h =  2n\int d^2 X W_h = 2\pi n \int_0^1 dh W_h = 2\pi n W(1).
\end{equation}
Obviously, the topological bound makes sense only provided that the ``superpotential" $W$ (the solution of the ``superpotential equation" (\ref{superpot-eq})) exists globally, i.e., in the full interval $h\in [0,1]$. The global existence of the superpotential is, therefore, a necessary condition for the existence of BPS soliton solutions. The issue of the global existence of the superpotential for a given potential is, in fact, quite nontrivial. But for the specific case of the old baby Skyrme potential in the BPS baby Skyrme model and, therefore, for the Higgs potential in the APD BPS vortex model, both a global superpotential and BPS soliton solutions do exist, for details we refer to \cite{bps-Sk1}. The same bound for the gauged BPS baby Skyrme model has already  been derived in \cite{stepien}. Finally we would like to add that   we have called $W$ the ``superpotential" and its defining equation (\ref{superpot-eq}) the ``superpotential equation", because they are exactly equal to the superpotential and superpotential equation which arises in supergravity coupled to a scalar field, see, e.g., \cite{f susy}.

\section{Summary}
In this paper we have introduced planar models with the area preserving diffeomorphisms symmetry of the static energy functional, which do support topological vortices, i.e., solitons carrying a topological charge $Q_v \in \pi_1(S^1)$. These solitons are BPS solutions, i.e., they solve a certain Bogomolny equation, and the corresponding total energy grows linearly with the topological charge. Due to a rather special form of the action (fixed by the symmetry requirement), which, among other things, does not contain the standard kinetic sigma model term, we could find vortices without having to introduce a gauge field. Moreover, perhaps because of the extremely large group of symmetries, we were able to find exact solutions for any topological charge in the case of a rather big family of Higgs type potentials with a U(1) vacuum manifold. (There is also an infinitely large group of target space symmetries which, being Noether symmetries, lead to infinitely many conserved charges. Concretely, these symmetries are given by an abelian subgroup of the area preserving diffeomorphisms oon the target space).
\\

The main result of this paper, however, is the observation that the APD BPS vortex model is, in fact, dual to the APD BPS baby Skyrme model. The duality is understood as the existence of a map (in this case a non-holomorphic map) between fields of these theories such that solutions of one model are transformed into solutions of the other one. Additionally, the energy densities of a vortex and of the corresponding baby skyrmion, as well as their charges, are exactly the same. In other words, the baby skyrmions of the APD BPS baby Skyrme model can be equivalently described in terms of vortices of the APD BPS vortex model. This is an extreme version of a previously found approximate correspondence between skyrmions (in the usual Skyrme models) and magnetic monopoles. Here, different topological objects are not only similar, but there are exactly equivalent to each other.         
\\

As the dual transformation holds also for time dependent solutions, one may easily show that these theories remain dual even at the semiclassical quantization level. Hence, the relevant excitation states should also be the same. 
\\

Finally, we have demonstrated that the same duality continues to hold for the gauged versions of these two models, where the basic complex scalar fields of the two models are coupled minimally to the electromagnetic field. As a consequence, the abelian Higgs model with the APD symmetry supports exactly the same BPS bound in terms of an auxiliary function (the superpotential) like the gauged baby Skyrme model with the same APD symmetry, and vortex solutions saturate this bound. We remark that other generalizations of the abelian Higgs model supporting BPS vortices have been introduced recently \cite{gen-ab-higgs}. 
\\

Undoubtedly, the most important question is whether the duality we have found  still exists in (3+1) dimensions, where now skyrmions of the VPD BPS Skyrme model would be dual to monopoles of some (currently unknown) VPD BPS monopole model. 
As already explained, the VPD BPS Skyrme model provides a useful starting point or a first approximation for the description of physical nuclei. 
This duality would, therefore, offer a dual description of baryons and atomic nuclei in terms of monopoles with an identification between the baryon charge and the monopole charge $Q_m \in \pi_2(S^2)$. Specifically, this would allow to identify monopole-like substructures within nuclei, which might provide valuable additional insight in their understanding. In this context we remark that in \cite{Sk monop} a version of the Yang-Mills-Higgs model was investigated, where a Skyrme type term quartic in covariant derivatives was included in addition to the standard term quadratic in covariant derivatives. The authors found that in this model the resulting ``skyrmed monopoles" behave differently from the skyrmions of the Skyrme model. Specifically their symmetries are different.

\section*{Acknowledgement}
The first three authors acknowledge financial support from the Ministry of Education, Culture and Sports, Spain (grant FPA2008-01177), 
the Xunta de Galicia (grant INCITE09.296.035PR and
Conselleria de Educacion), the
Spanish Consolider-Ingenio 2010 Programme CPAN (CSD2007-00042), and FEDER. 
Further, AW was supported by Polish NCN grant 2011/01/B/ST2/00464.

\end{document}